\documentclass[prl,floatfix,showpacs,superscriptaddress,amsmath,twocolumn]{revtex4}

\usepackage{epsfig,amssymb,amsfonts}
\usepackage{amsmath}
\usepackage{verbatim}

\begin{document}

\title{Noise-Free Measurement of Harmonic Oscillators with Instantaneous Interactions}

\date{\today}

\author{M. Fran\c{c}a Santos}\email{msantos@fisica.ufmg.br}

\affiliation{Dep. de F\'{\i}sica, Universidade Federal de Minas
Gerais, Belo Horizonte, 30161-970, MG, Brazil}

\author{G. Giedke}\email{giedke@mpq.mpg.de}

\affiliation{Max-Planck-Institut f\"ur Quantenoptik,
Hans-Kopfermann-Str. 1, D-85748 Garching, Germany}

\author{E. Solano}\email{enrique.solano@physik.lmu.de}

\affiliation{Max-Planck-Institut f\"ur Quantenoptik,
Hans-Kopfermann-Str. 1, D-85748 Garching, Germany}

\affiliation{Physics Department, ASC, and CeNS,
Ludwig-Maximilians-Universit\"at, Theresienstrasse 37, 80333 Munich,
Germany}

\affiliation{Secci\'on F\'{\i}sica, Departamento de Ciencias,
Pontificia Universidad Cat\'olica del Per\'u, Apartado Postal 1761,
Lima, Peru}

\begin{abstract}
We present a method of measuring the quantum state of a harmonic
oscillator through instantaneous probe-system selective interactions
of the Jaynes-Cummings type. We prove that this scheme is robust to
general decoherence mechanisms, allowing the possibility of
measuring fast-decaying systems in the weak-coupling regime. This
method could be applied to different setups: motional states of
trapped ions, microwave fields in cavity/circuit QED, and even intra-cavity
optical fields.
\end{abstract}


\maketitle

Measuring the quantum state of a harmonic oscillator, or,
equivalently, its associated Wigner function~\cite{Glauber}, is a
 fundamental task of quantum physics. Some
proposals allow a direct measurement, like propagating optical
fields tested with homodyning techniques~\cite{Leonhardt}. Others,
due to problems of accessibility, require indirect measurement
schemes via interaction with a probe. This is the case of
microwave fields in 3D cavities~\cite{ReviewParis}, circuit cavity
QED with superconducting qubits~\cite{Yale}, or the motion of
trapped ions~\cite{ReviewIons}. Different as they are, known
techniques share a common problem: the noisy action of decoherence
due to the probe-system {\it finite interaction times}.

Recently, the reconstruction of a Wigner function in microwave
cavity QED (CQED) was successfully realized~\cite{PhotonWigner} with
the aid of a dispersive probe-system interaction~\cite{Lutter}.
Unfortunately, dispersive coupling is known to be slow and the
required interaction time allows decoherence processes to disturb
the measurement. In Ref.~\cite{FresnelWigner}, a resonant method was
proposed, but the Wigner reconstruction depends on the possibility
of monitoring a few Rabi cycles, adding up to long probe observation
periods. In the case of trapped ions, the measurement techniques are
quite similar and a recent experiment~\cite{PhononWigner} made use
of numerical integration over several Rabi cycles to achieve the
goal. These long interaction times are particularly harmful in the
case of fast-decaying systems. For example, state reconstruction of
intracavity optical fields has not been experimentally attempted, to
our knowledge, due to their weak coupling with atomic probes.

In this paper, we present a method to measure the quantum state of a
harmonic oscillator through {\it instantaneous} probe-system
interactions~\cite{AtomHomodyning,IonHomodyning}, preventing
decoherence from disturbing the measurement. The harmonic oscillator
is allowed to interact with a two-level probe for an arbitrarily
short time via a selective
interaction~\cite{CQEDSelectivity,IonSelectivity,UnivSelectivity} in
the Jaynes-Cummings (JC) model~\cite{JC}. The information is then
collected from the second time derivative of the probe population at
{\it zero interaction time}. The scheme permits to measure the
population field distribution and, with the support of coherent
displacements, the associated Wigner and $Q$ functions at any point
in phase space with arbitrarily small influence of decoherence. From
this data the full Wigner function can be reconstructed by a simple
fit as in \cite{PhotonWigner} or by more sophisticated techniques
\cite{StateEstimation1}, e.g., involving maximum likelihood
estimation~\cite{StateEstimation2} and taking into account the imperfections
of the measurement process \cite{Leonhardt}.

Typically, a selective interaction can be built when a three-level
probe, driven by a classical and a quantized field, is reduced to
two metastable states after adiabatic elimination of the third
level, allowed by a large detuning~$\Delta$. The associated
Hamiltonian reads~\cite{CQEDSelectivity,IonSelectivity}
\begin{eqnarray}
\label{SelHam1} \hat{H}_{\rm eff } = \!\!\!\! && \hbar
\frac{\Omega_{1}^2}{\Delta}
 | g \rangle \langle g | + \hbar \frac{\Omega_{2}^2}{\Delta}
\hat{a}^\dagger\hat{a} | e \rangle \langle e | \nonumber \\ && +
\hbar \frac{\Omega_{1} \Omega_{2}}{\Delta}\big( | g \rangle
\langle e | \hat{a}^\dagger + | e \rangle \langle g | \hat{a}
\big) .
\end{eqnarray}
Here, $\{ |g \rangle , | e \rangle \}$ are the (metastable) ground
and excited states of the two-level probe, $\{ \hat{a} ,
\hat{a}^{\dagger} \}$ are the harmonic oscillator annihilation and
creation operators, respectively, and $\Omega \equiv \Omega_{1}
\Omega_{2} / \Delta$ is the effective JC coupling strength. The
first and second terms on the r.h.s. of Eq.~(\ref{SelHam1}) are
AC-Stark shifts associated with each of the probe levels, the
second one depending on the number of oscillator excitations. This
means that any effort at tuning the JC coupling to resonance will
succeed only for a selected JC doublet ${\cal H}_{N}: \{ | g
\rangle | N \rangle , | e \rangle | N-1 \rangle \}$, leaving all
other doublets, for which $n \neq N$, slightly or completely
off-resonance. It can be shown that, under proper tuning of the
excitation fields, the condition $\Omega_{2} \gg \Omega_{1}
\sqrt{N}$ assures neat selectivity in the JC model, where resonant
Rabi oscillations will happen only inside the subspace ${\cal
H}_{N}$. In this case, Eq.~(\ref{SelHam1}) turns into the
selective Hamiltonian
\begin{eqnarray}
\label{SelHam2}  \widehat{H}_{N} \! = \!  \hbar \sqrt{N} \Omega \big( | g
\rangle \langle e |  \otimes  | N \rangle \langle N-1 | + | e
\rangle \langle g | \otimes | N-1 \rangle \langle N | \big) , \nonumber \\ 
\end{eqnarray}
describing the flip-flop interaction of two effective spin-$1/2$
systems, $\{ |g \rangle , | e \rangle \}$ and $\{ | N-1 \rangle ,
| N \rangle \}$. Specific implementations have been proposed in
microwave CQED~\cite{CQEDSelectivity} and in trapped
ions~\cite{IonSelectivity}, but other systems, like an atom inside
an optical cavity, or a superconducting qubit coupled to a
coplanar waveguide resonator, can also enjoy a similar behavior.

From the unitary time evolution of the total density operator,
$\dot{\hat \rho} = [ {\widehat{H}}_N , {\hat \rho} ] / i \hbar$,
the first and second time derivatives of the expectation value of
a time-independent probe operator $\hat B$ can be expressed as
\begin{eqnarray}
\label{Ehrenfest1} \frac{d \langle \hat{B} \rangle}{dt} & = &
\frac{1}{i \hbar} \langle [ {\hat B} , {\widehat H}_N ] \rangle  ,
\\ \label{Ehrenfest2} \frac{d^2 \langle \hat{B} \rangle}{dt^2} & =
& \frac{1}{(i \hbar)^2} \langle \big[ [ {\hat B} , {\hat H}_N ] ,
{\hat H}_N \big] \rangle .
\end{eqnarray}
We study here a more general case, allowing decoherence in the
(field) system but disregarding decoherence in the probe. We do
that based on the fact that most physical setups use probes with
long lifetimes compared to the ones of the systems to measure.
Under this assumption, we consider the most general master
equation $\dot{\hat {\rho}} = {\cal L} \hat\rho$ in the Lindblad
form~\cite{breuer},
\begin{eqnarray}
\label{lindblad} \dot{\hat\rho} \!\! & = & \!\!\! \frac{1}{i \hbar}
\left[ \hat{H}_N , \hat\rho \right] \nonumber \\
&& + \! \sum_m \! \kappa_m ( \hat{A}_m \hat\rho
\hat{A}_m^{\dagger} - \! \frac{1}{2} \hat{A}_m^{\dagger} \hat{A}_m
\hat\rho - \! \frac{1}{2} \hat\rho \hat{A}_m^\dagger \hat{A}_m )
\, ,
\end{eqnarray}
where $\kappa_m$ express decay rates and Lindblad operators $A_m$
and $A_m^{\dagger}$ are associated with the (field) system.

We calculate first
\begin{eqnarray*}
\frac{d \langle \hat{B} \rangle}{dt} =
\!\!\!\!\!\!\! &&
 \frac{1}{i \hbar} \langle [ {\hat B} , {\hat H}_N ]
\rangle \nonumber \\ && +
{\rm Tr} \bigg\lbrack \! \sum_m \!
\kappa_m ( \hat{A}_m \hat\rho \hat{A}_m^{\dagger} \! - \!
\frac{1}{2} \hat{A}_m^{\dagger} \hat{A}_m \hat\rho - \!\!
\frac{1}{2} \hat\rho
\hat{A}_m^\dagger \hat{A}_m ) \hat{B} \bigg\rbrack . \nonumber
\end{eqnarray*}
The second term on the r.h.s. can be rewritten under the trace as
\[
{\rm Tr} \frac{1}{2}\bigg\lbrack \! \sum_m \! \kappa_m
\hat\rho \left( [\hat{A}_m^\dagger,\hat{B}]\hat{A}_m + \hat{A}_m^{\dagger}
[\hat{B},\hat{A}_m] \right)\bigg\rbrack
\]
and, since  $[ {\hat B}, {\hat A}_m^{\dagger} ] = [ {\hat B}, {\hat
A}_m ] = 0$, we have
\begin{eqnarray}
\label{DissEhrenfest1} \frac{d \langle \hat{B} \rangle}{dt}
& = &  \frac{1}{i \hbar} \langle [ {\hat B} , {\hat H}_N
] \rangle.
\end{eqnarray}

In this expression, see similarity with Eq.~(\ref{Ehrenfest1}), the
dynamics of probe operator expectation value $\langle {\hat B} \rangle$ does not seem
affected by the system decay. This is certainly not the case, as the
time-dependent expectation value on the r.h.s. of
Eq.~(\ref{DissEhrenfest1}) will in general be susceptible to
decoherence: the calculation involves time-dependent $\rho(t)$
following Eq.~(\ref{lindblad}). However, at $t=0$ the noise terms
vanish identically, i.e., the time derivative of any probe
expectation value at $t=0$ is independent of any field-decohering
Lindblad environment. For this particular time, choosing a probe
operator ${\hat B} = | e \rangle \langle e |$ and probe-system
initial state $\rho (0) = | g \rangle \langle g | \otimes \rho_f$,
we obtain
\begin{eqnarray}
\label{Ehrenfest2zero} \frac{d P_e (\tau)}{d{\tau}} \bigg|_{\tau =
0} = 0 ,
\end{eqnarray}
with dimensionless time $\tau \equiv \Omega t$ and $P_e(\tau)
\equiv \langle | e \rangle \langle e | \rangle$.
For the second derivative, we obtain
\begin{eqnarray}
\label{PrePN} \frac{d^2 \langle \hat{B} \rangle}{dt^2} =
\!\!\!\!\!\!\! && \frac{1}{(i \hbar)^2} \langle \big[ [ {\hat B} ,
{\hat H}_N ] , {\hat H}_N \big] \rangle + {\rm Tr} \bigg\lbrack
\sum_m
\kappa_m ( \hat{A}_m \hat\rho \hat{A}_m^{\dagger} \nonumber \\
&& - \frac{1}{2} \hat{A}_m^{\dagger} \hat{A}_m \hat\rho -
\frac{1}{2} \hat\rho \hat{A}_m^\dagger \hat{A}_m ) [ {\hat
B} , {\hat H}_N ] \bigg\rbrack  .
\end{eqnarray}
This is a different situation and the trace will not vanish in general, as
before. However, considering again the observable ${\hat B} = | e
\rangle \langle e |$ and probe-system initial state $\rho (0) = | g \rangle
\langle g | \otimes \rho_f$, it follows that the second term in
Eq.~(\ref{PrePN}) vanishes again, and we have
\begin{eqnarray}
\label{PN} \frac{d^2 P_e ( \tau )}{d{\tau}^2} \bigg\vert_{\tau = 0}
= P_N .
\end{eqnarray}
Here, $P_N = {\rm Tr} \lbrack \rho_f | N \rangle \langle N | \rbrack$
is the probability of finding Fock state $| N \rangle$ in the
initially unknown harmonic oscillator state.

Equation~(\ref{PN}) describes a remarkable result, it shows that the
curvature of the function $P_e (\tau)$, at vanishing $\tau = 0$,
contains {\it undisturbed} information about $P_N$. This valuable
field information is encoded correctly in the level statistics of
the two-level probe even in presence of a field reservoir of a
general kind. Needless to say, all previous results hold when the
usual thermal bath is considered as a reservoir for the harmonic
oscillator. At zero temperature, for example, Lindblad operators
$\hat{A}_m$ and $\hat{A}_m^{\dagger}$ would have to be replaced by
$\hat{a}$ and $\hat{a}^{\dagger}$, respectively, and $\kappa$ would
represent the decay rate of the single field mode. The
counter-intuitive results of (\ref{DissEhrenfest1}) and (\ref{PN})
are of an infinitesimal nature and, without harming their
theoretical importance, should suffer high-order corrections when
dealing with a discrete sampling of interaction times, as will be
explained later.

In order to measure the complete field population, $P_n$, $\forall
n$, one just needs to tune resonantly the other selected subspaces
${\cal H}_n$ and follow a similar procedure.  The measurement of all
$P_n$ allows the estimation of the complete Wigner function
$W(\alpha)$ of $\rho$, conditioned to the realization of previous
arbitrary field displacements $D(-\alpha)$ in phase space. For that,
we have to recall that the Wigner function can be
expressed~\cite{Glauber}, among other possibilities, as
\begin{eqnarray}
\label{Wignerfunction} W( \alpha) \equiv
2\sum_{n=0}^{\infty}(-1)^nP_n(-\alpha) ,
\end{eqnarray}
where $P_n(-\alpha)$ stands for the field population after a
displacement $D(-\alpha)$. Another rather original manner of
reconstructing the quantum field state in phase space is via the
instantaneous measurement of the $Q$-function~\cite{Glauber},
defined as $Q = \langle \alpha | \rho_f | \alpha \rangle$. It can be
shown that
\begin{eqnarray}
\!\!\!\!\! Q (\alpha) = && \!\!\!\!\!\! {\rm Tr} \lbrack D ( -\alpha
) \rho_f D^{\dagger} ( -\alpha) | 0 \rangle \langle 0 | \rbrack =
{\rm Tr} \lbrack \rho_f (
-\alpha ) | 0 \rangle \langle 0 | \rbrack \nonumber \\
= && \!\!\!\!\!\! P_0 (-\alpha) .
\end{eqnarray}
This means that, following Eq.~(\ref{PN}), measuring instantaneously
the probability of having Fock state $| 0 \rangle$ after field
displacements $D( -\alpha )$, amounts to a full measurement of the
$Q$-function. Note that this will only require the tuning of a
single selective subspace, reducing enormously the experimental
efforts when compared to the case of the Wigner function. Typically,
coherent displacements can be realized at a very high rate,
depending mainly on the intensity of the excitation fields, so this
is not a critical issue.

From a fundamental point of view, our proposal suggests that, no
matter how short the lifetime of a certain system is, there would always
exist the possibility of encoding its quantum information in
two-level probe statistics at infinitesimal interaction times. In
other words, measuring quantum states may not require long-living
systems or strong probe-system coupling, against common belief. In
this work, we have not proved this conjecture in general. However,
we have given a particular example, the case of a quantum harmonic
oscillator or a single mode field, that is applicable to many
physical setups.

Turning to more practical considerations, we stress that our scheme
employs exclusively, as a final readout mechanism, the measurement
of the population of the excited state of a two-level probe at
different times. This probe population is measured directly by ionization, in the case of cavity QED, or fluorescence, for trapped ions and atoms, with suitable additional fields, different from the ones used in Eq. (1). This is a standard measurement in many
quantum-optical experimental setups of interest and has been
realized routinely with high accuracy and efficiency
\cite{PhotonWigner,PhononWigner,PeMeasurement}.

Estimating derivatives from a finite-time experimental sampling will
produce higher order corrections and an overhead in the number of
repetition measurements. In consequence, the benefit of
instantaneous measurements demands an improved measurement accuracy,
which can be seen as follows. To determine the discrete second
derivative in Eq.~(\ref{PN}), $\ddot{P}_e (0)$, we measure $P_e$ at
$0, \tau, 2\tau$, for small $\tau$, and calculate
\begin{eqnarray}
\label{discretederivative} \frac{P_e(2\tau)-2P_e(\tau)+P_e(0)}{
\tau^2} = \ddot{P}_e (0) + o(\tau),
\end{eqnarray}
The measurement results are scattered around each $P_e$ with
variance $\Delta^2_e = \Delta_q^2+\Delta_t^2$, referring to
quantum-mechanical and technical noise, the latter arising from
imperfections in preparation, timing, and measurement. For small
values of $\tau$ and probe operator
$\hat{B} = |e\rangle
\langle e |$, we have $\Delta^2_q(\tau) = \langle \hat{B}^2
\rangle - \langle \hat{B} \rangle^2 = P_e(\tau)-P_e(\tau)^2
\sim P_N\tau^2$, following the Taylor expansion of $P_e(\tau)$
around zero time. Performing $M$ measurements at each of the three
times of Eq.~(\ref{discretederivative}), we can reduce the
uncertainty in the three expectation values by a factor
$1/\sqrt{M}$. Thus, our estimate for the second derivative comes
with an error variance
\begin{equation}
\Delta^2 \sim \frac{6P_N\tau^2+ 4\Delta^2_t}{M\tau^4} ,
\end{equation}
while the {\it signal} is $\sim P_N$. Then, in order to achieve a
signal-to-noise ratio larger than unity, $P_N / \Delta > 1$, we
need a number of measurements
\begin{equation}
M > \tau^{-2}+\Delta_t^2\tau^{-4} ,
\end{equation}
approximately. In this way, as we reduce $\tau$ by a factor $f$, in
order to improve the approximation of $\ddot{P}_e (0)$, we need to
increase the number of measurements by $f^2$ (or $f^4$  if technical
errors dominate) to maintain the desired signal-to-noise ratio.
Consequently,  the total probe-system interaction time $M\tau$
summed over all measurements increases, while each measurement
outcome represents the effect of an arbitrarily short interaction
time $\tau$. Although this is a high price to pay, contamination of
the oscillator by decoherence can be kept arbitrarily small, no
matter how fast it is.

Quantitatively, the determination of $\ddot{P_e}(0)$ is associated
with an error $\sim \dddot{P_e}(0)\tau$ due to the finiteness of
$\tau$. In this case, we can use Eq.~(\ref{lindblad}) to obtain an
expression for $\dddot{P_e}(0)$ and observe two contributions: a
unitary component, present even in the absence of noise, and an
additional term $\propto\kappa/\Omega$ due to the noise
\footnote{Possible terms
$\propto(\kappa/\Omega)^2,(\kappa/\Omega)^3$ can be seen to
vanish.}.  In order to determine $\ddot{P_e} (0)$ to accuracy $q$,
we need $\tau<q/\dddot{P_e} (0)$.  This, in turn, requires $M
>q^{-2}\dddot{P_e}(0)^2$ for ideal measurements and $M >
q^{-4}\dddot{P_e}(0)^4\Delta_t^2 $ in the case of
technical-dominated errors.  Hence, the number $M$ of required
measurements increases polynomially with $\kappa$ (in the case
$\kappa\gg\Omega$ like $\kappa^2$ resp. $\kappa^4$). Remarkably, no
matter how strong $\kappa$ is, we can always make sure that it does
not affect the measurement result by shortening $\tau$.
This procedure finds its natural limit $\tau > \Omega / \omega$, where
$\omega$ is the smallest dominant frequency of the specified dynamics
($\omega = \Omega_2^2 / \Delta$ in our example, which implies
$\tau>\Omega_1/\Omega_2$). Note that we only need
to know an upper bound of the decoherence rate to perform an
accurate measurement, while alternative methods involve specific
decoherence models.

The proposed method can be applied to any physical system enjoying
JC selective interactions, and may allow for the complete
measurement of elusive fast-decaying systems, as is the case of
intracavity optical fields, among others. Furthermore, since the
scheme is based on infinitesimal transfer of information, this
method could observe initial state conditions. For example, it
could be used to test the initial purity of a system, as well as
its unavoidable entanglement with external degrees of freedom at
zero interaction time. This may prove useful for testing the
validity of usual assumptions taken in decoherence theories, such
as system and environment being initially in a separable state or
markovian dynamics.

\begin{acknowledgments}

The authors thank Benasque Quantum Information Workshop 2005 for
sporting and also for scientific hospitality. E.S. acknowledges
support from DFG SFB 631, EU RESQ and EuroSQIP projects. M.F.S. received financial support from Infoquant/CNPq.

\end{acknowledgments}

\end{document}